# Parametric Tracking of Electrical Current Components Using Gradient Descent Algorithm

Marouane Frini[1], Vincent Choqueuse[2] and François Auger[1]

[1] IREENA, Université de Nantes, BP 406, 44602 Saint-Nazaire cedex, FRANCE.

[2] LAB-STICC, Université de Bretagne Occidentale, 29238 Brest Cedex 3, FRANCE

**ABSTRACT**

In the last few years, Motor Current Signature Analysis (MCSA) has proven to be an effective method for electrical machines condition monitoring. Indeed, compared to vibration and temperature analysis, current measurement proves to be a convenient and non-invasive alternative. Moreover, it has proven to be a reliable method since many mechanical and electrical faults manifest as side-band spectral components generated around the fundamental frequency component of the motor's current. These components are called interharmonics and they are a major focus of fault detection using MCSA. However, the main drawback of this approach is that the interference of other more prevalent components such as the fundamental and noise components can obstruct the effect of interharmonics in the spectrum and may therefore affect fault detection accuracy.

Thus, we propose in this paper an alternative approach that aims to decompose the different current components using a model based on a Vandermonde matrix, in order to monitor each component independently. Then, the tracking of each distinct component in time and spectral domains is implemented. This is achieved by estimating their respective relevant parameters using the Gradient Descent algorithm. This method has been favorably compared to an existing estimation algorithm (MUSIC) and its efficiency has been validated. The results of this work prove to be promising and establish the parametric tracking of the electrical current components using the Gradient Descent algorithm as a reliable monitoring approach.

# 1. INTRODUCTION

In the context of condition monitoring techniques for predictive maintenance, there is a constant search for improvements in the measurement process in order to facilitate the technical interventions and reduce maintenance downtime. Lately, Motor Current Signature Analysis (MCSA) has been rapidly gaining a wide acceptance in many industrial applications thanks to its non-invasiveness, its ease of implementation and its overall low-cost [1].

In contrast to classical methods such as vibration and temperature analysis, MCSA only requires the motor's electrical measurements that are often already monitored for machine protection and are therefore easily accessible. It has been proven that any mechanical (bearing damage, gear wear, shaft eccentricity…) and electrical (phase unbalance, power surges…) fault that appears across any element of the transmission system is bound to induce a shift in the rotating flux components of the induction motor [2,3]. Specifically, these faults cause a magnetic field disturbance, thus changing the mutual and self-inductances of the electric motor, leading to the creation of side-bands across the main frequency component spectrum [4]. These fault related components are commonly referred to as interharmonics, since they appear between the fundamental frequency component and the harmonic frequencies component [5].

Considering that the main goal of the MCSA is the fault monitoring from its early stages of development, great emphasis is placed on the accurate detection of these interharmonics as soon as possible [6]. Several works were based on MCSA using periodograms, in order to evaluate the spectral density of the current signal related to different fault types [7, 8].

However, these techniques are limited by their restricting spectral resolution. This implies that the interharmonic components can be obfuscated by the more pervasive neighbouring supply frequency dynamics and by the noise influence, therefore hindering the fault detection process [9].

There are several techniques in literature that have been used to decompose the current signals for fault detection such as Wavelet Analysis and Empirical Mode Decomposition (EMD) [5]. However, these techniques are not physically suitable in regards to the electrical currents components and they are mainly limited by their computational intensiveness and the loss of the original signal information quantity.

Thus, we propose in this paper a more natural and convenient approach that aims to use a model based on the Vandermonde matrix of the current signal, in order to extract its various components: the fundamental, the harmonics, the interharmonics and the residual components.

Moreover, the parametric tracking of each component in time and spectral domains based on the Gradient Descent algorithm's estimation of their relevant parameters is implemented. This establishes the condition monitoring of electrical current through the parametric tracking of its components.

Therefore, the organization of this paper in the subsequent sections is as follows. In section 2, the theory of the proposed current signal model based on the Vandermonde matrix is detailed. Next, section 3 presents the Gradient Descent algorithm used for parameters estimation. The estimation results are shown in section 4 and they are validated through the comparison to an existing method (MUSIC) in section 5. Finally, the last section provides some global conclusions.

## 2. SIGNAL MODEL

Even though the motor current signal can be ideally represented as a simple sinus wave with a given supply frequency $f_0$, in real conditions this signal can contain additional spectral components. These components range from harmonic components, which are non-linear elements coming from in the power supply's load, to interharmonic components which are related to mechanical or electric faults as well as noise elements introduced by various sources [10].

Thus, the electrical current signal can be represented as a sum of sinusoids with an added noise component. In complex form, it can be written as:

$$y[n] = \sum_{l=1}^{L} c_l e^{j\omega_l n} + b[n] \qquad (1)$$

where $L$ refers to the number of spectral components of the signal, $c_l$ represents the complex phasors, $\omega_l = 2\pi f_l / f_s$ represents the angular frequencies which are normalized by the sampling frequency $f_s$, $n$ represents the time index and $b[n]$ refers to the noise component. In matrix form, the current signal $y$ can be written as:

$$\boldsymbol{y} = \mathbf{V}(\boldsymbol{\omega})\mathbf{c} + \mathbf{b} \qquad (2)$$

where $\mathbf{V}(\omega)$ is an $N$ x $L$ Vandermonde matrix [11]. It is defined as:

$$\mathbf{V}(\omega) = \begin{bmatrix} 1 & \cdots & 1 \\ e^{j\omega_1} & & e^{j\omega_L} \\ \vdots & \ddots & \vdots \\ e^{j\omega_1(N-1)} & \cdots & e^{j\omega_L(N-1)} \end{bmatrix} \quad (3)$$

where $\omega = [\omega_1, \ldots, \omega_L]$ represents the line vector containing $L$ normalized angular frequencies, $c = [c_1, \ldots, c_L]^T$ represents the $L$ sized column vector containing the complex phasors and $b = [b[0], \ldots, b[N-1]]^T$ represents the $N$ sized column vector containing the noise samples.

The model based on the Vandermonde matrix has been used for a general $\omega$ and $c$ parametric estimation [11]. However, in the case of current signals, different physical phenomena tend to introduce spectral components with specific $\omega$ and $c$ signatures. Therefore, the idea of independently estimating these specific parameters can notably improve estimation performance. In the case of electrical machines, the spectral components can be regrouped in the following categories according to their physical origin:

- The fundamental component: in an ideal electrical motor, the stator current can be represented by a sinusoid with a fundamental angular frequency $\omega_0$ which is imposed by the supply network. Its associated ordinary frequency value is fixed to 50 Hz or 60 Hz according to the supply network geographic location.

- The harmonic components: these specific components are introduced by the non-linear loads in the power supply grid. Since these load components are generally symmetrical, harmonics have an angular frequency $\omega_h$ that is a positive integer multiple of the angular frequency of the fundamental component and they are expressed as:

$$\omega_h = h\,\omega_0 \quad (4)$$

with $h \in \mathbb{Z}^*$.

- The interharmonic components: these components are introduced by the behavioural modification of the motor's electromagnetic field due to mechanical and electrical faults. These faults give rise to additional components as:

$$\omega_i = \omega_0 \pm k\,\omega_c \quad (5)$$

with $k \in \mathbb{Z}^*$ and $\omega_c$ represents the characteristic angular frequency of the fault which depends on the fault type [12].

Hence, we establish the decomposition of the current signal based on the different components presented in (4) and (5) using the aforementioned model expressed in (2) and (3). Consequently, the current signal can be represented in complex form as follows.

$$y[n] = y_f[n] + y_h[n] + y_i[n] + b[n] \qquad (6)$$

where:

- $y_f[n] = c_1 e^{j\omega_0 n}$ represents the fundamental component.
- $y_h[n] = \sum_{l=2}^{L} c_l e^{j\omega_0 l n}$ represents the harmonic component.
- $y_i[n] = \sum_{k=0}^{K} c_k e^{j(\omega_0 + k\omega_c)n}$ represents the interharmonic component.
- $b[n]$ represents the residual component.

Therefore, in order to establish the efficient condition monitoring of the electrical current and enable improved fault detection, the fundamental, the harmonic, the interharmonic and the residual components should be separately represented in time and frequency domains. To achieve this goal, the tracking of the associated parameters evolution must be implemented. These parameters are the fundamental angular frequency $\omega_0$, the harmonic phasor $c_l$, the interharmonic angular frequency $\omega_c$ and the interharmonic phasor $c_k$.

The estimation of the aforementioned parameters is implemented using an optimization algorithm based on the Gradient Descent approach. This algorithm is detailed in the following section.

## 3. ESTIMATION ALGORITHM

The aim of the estimation algorithm presented in this section is to track the previously established parameters $c_l$, $c_k$, $\omega_0$ and $\omega_c$ in the context of an optimization problem. This ensures the accurate prediction of the proposed signal model.

Hence, the main objective is to estimate the four optimal parameters so that they minimize a cost function representing the difference between the original current signal and the previously established model based on Vandermonde matrix.

If we consider that the original current signal with n data points is subdivided into several consecutive segments *M* of samples without overlap between segments, then each segment of this signal can be represented as:

$$y_n = [y[nM], y[nM+1], \ldots, y[nM+M-1]]^T \tag{7}$$

Thus, the estimation algorithm aims to update the parameters $c_l$, $c_k$, $\omega_0$ and $\omega_c$ by minimizing the cost function (a convex function having a global minimum) involving the difference between the original signal *y[n]* as expressed in (6) and the model as established in (2). This cost function is written as:

$$J_n(\omega_0, \omega_c, C) = \|y_n - V(\omega_0, \omega_c)C\|^2 \tag{8}$$

where the matrix C contains the $c_u$ phasors for $u \in \mathcal{K} \cup \mathcal{I}$ with $k \in \mathcal{K}$ and $l \in \mathcal{I}$, and $V(\omega_0, \omega_c)$ is the Vandermonde Matrix depending on $\omega_0$ and $\omega_c$.

In order to find the different parameters that minimize the cost function (8), the Gradient Descent algorithm [13] is chosen as an optimization method to ensure that the proposed model makes accurate predictions. This algorithm has been gaining a rapidly increasing attention for machine learning and deep applications, especially for neural networks weights optimization. Compared to the Levenberg–Marquardt Algorithm (LMA) which can be theoretically considered as a more appropriate method in the case of a limited number of parameters, the Gradient Descent has a slower overall convergence rate but it is more robust in regards to the initialization choice [13]. Furthermore, in this case, the Gradient Descent algorithm represents a more intuitive, less computationally demanding and less complex alternative to the LMA [14].

Basically, the Gradient Descent algorithm's prediction is based on first order simple linear regression. In other terms, it aims to find a prediction function associated to an input predictive variable. This function represents the most optimal approximation that minimizes the global error.

Hence, the goal of the simple linear regression is to ultimately find the optimal predictive coefficients so that the prediction function is the closest possible to the estimated variable for every pair of variables forming the learning data set. In order to find the best parameters, the associated cost function based on the mean squared error needs to be minimized [15].

In the context of the proposed parameters estimation, the Gradient Descent algorithm finds the optimal parameters $\omega_0$, $\omega_c$, $c_l$ and $c_k$ that satisfy the minimization criteria of the cost function $J_n(\omega_0, \omega_c, C)$ as expressed in (8). The main steps of this algorithm are presented as follows.

- Step 1: Initialize the values of the parameters $\omega_0$, $\omega_c$, $c_l$ and $c_k$.
- Step 2: Repeat the following mathematical operations until reaching convergence to the global minimum of the cost function.

$$\omega_0^{i+1} = \omega_0^i - \mu \frac{\partial J_n}{\partial \omega_0}(\omega_0, \omega_c, C) \quad (10)$$

$$\omega_c^{i+1} = \omega_c^i - \mu \frac{\partial J_n}{\partial \omega_c}(\omega_0, \omega_c, C)$$

$$c_l^{i+1} = c_l^i - \mu \frac{\partial J_n}{\partial c_l}(\omega_0, \omega_c, C)$$

$$c_k^{i+1} = c_k^i - \mu \frac{\partial J_n}{\partial c_k}(\omega_0, \omega_c, C)$$

- Step 3: Return the updated parameters $\omega_0$, $\omega_c$, $c_l$ and $c_k$.

The coefficient µ represents the learning rate of the Gradient Descent algorithm. It is a tuning parameter that dictates the step size at each iteration while moving toward the minimum of the cost function. It is chosen empirically so that it satisfies a tradeoff between a high and a low value. Indeed, if the learning rate µ is too high the algorithm will not converge since it will overshoot and it will oscillate around the desired minimum without reaching it. However, if the learning rate is too small, the algorithm's descent will be too slow and it will not converge given the limited number of iterations. The choice of the iteration number I is also an important parameter in order to efficiently reach convergence [15]. The following section presents the results of this estimation algorithm.

4. **RESULTS**

This section details the results of the previously presented estimation algorithm and the tracking of the derived parameters $\omega_0$, $\omega_c$, $c_l$ and $c_k$ involving a reference current signal and the Vandermonde model as shown in (8).

The reference current signal $y_n$ is a synthetic complex signal characterized by a sampling frequency $f_s$ = 1000 Hz, a signal duration $T$ = 1 s and a data point number $N$ = 1000 points.

The fundamental frequency is $f_0$ = 60 Hz, the fundamental amplitude is $A_0$ = 0.7 A and the fundamental phase is $\varphi_0$ = 0 rad.

The number of harmonics is $l$ = 3, their respective amplitudes are $A_l$ = [0.6 0.5 0.4] A and their phases are $\varphi_l$ = [0 0 0] rad.

The number of interharmonics is $k$ = 3, the fault characteristic frequency is $f_c$ = 5 Hz, the harmonics amplitude is $A_k$ = [0.3 0.2 0.1] A and their phases are $\varphi_k$ = [0 0 0] rad.

The introduced noise is a Gaussian White Noise with a mean value $\mu$ = 0 and a standard deviation $\sigma$ = 0.25.

In the case of real signals, the conversion to complex form can be achieved by filtering out the negative-frequency component of the signal. This conversion is done by using a Hilbert Transform filter (large order FIR filter) to find the imaginary part [16].

The resulting signal $y_n$ in spectral domains are shown in Fig. 1.

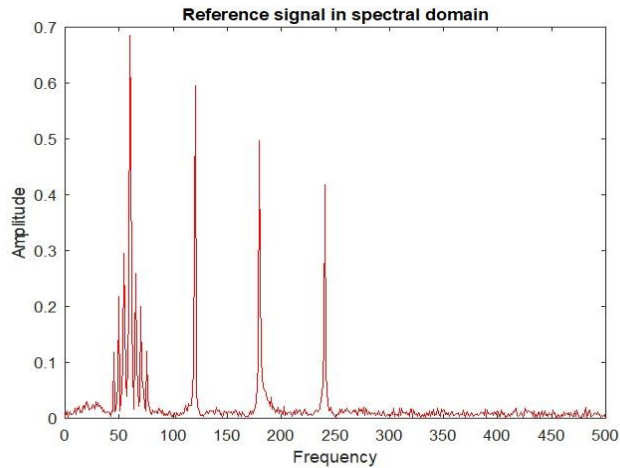

Figure 1 : Reference signal $y_n$ in spectral domain.

The signal generated based on the proposed model $V(\omega_0, \omega_c)C$ according to Section 2 has the same characteristics of the previously presented reference signal in terms of components numbers and values.

Thus, the real part of this modeled signal $V(\omega_0, \omega_c)C$ in spectral domains are shown in Fig. 2.

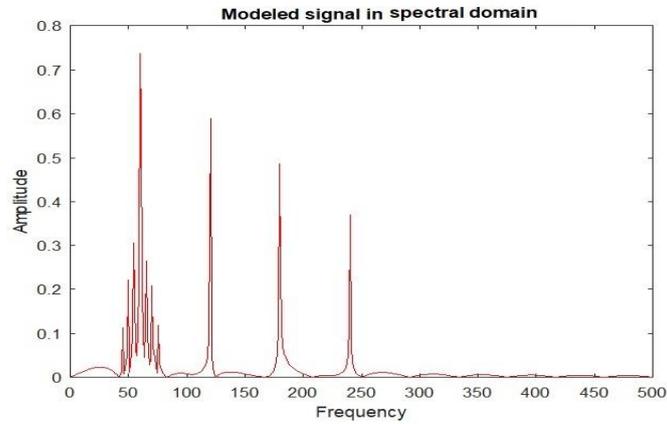

Figure 2: Modeled signal $V(\omega_0, \omega_c)C$ in spectral domain.

It can be clearly seen from Fig. 1 and Fig. 2 that both the reference signal $y_n$ and the modeled signal $V(\omega_0, \omega_c)C$ contain all the expected current components at the appropriate frequencies in accordance to the current signal theory presented in Section 2.

The signal is subdivided into 4 segments as expressed in (7) with M = 250 data points and they are introduced as inputs in the implemented estimation algorithm as shown in (12). According to the explanation in Section 3, the chosen learning rate is µ = 0.004 and the number of iterations is $i$ = 250 in order to reach convergence with the best possible efficiency. The parameters $\omega_0$, $\omega_c$, $c_l$ and $c_k$ are initialized to zero. The evolution of the updated cost function $J_n$ according to the iterations number $I$ resulting from the algorithm's estimation is shown in Fig 3.

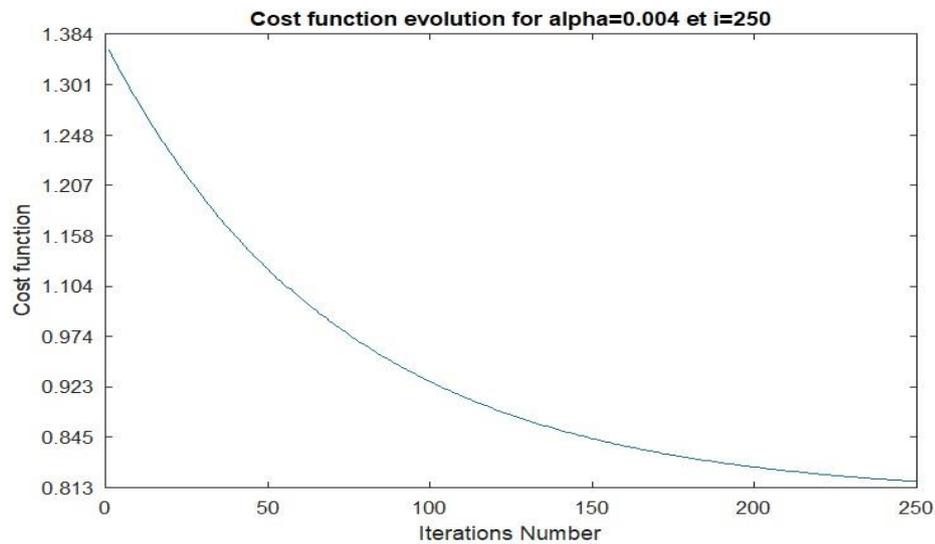

Figure 3: Cost function evolution for $\mu$ = 0.004 and I =250.

The evolution of the cost function in Fig. 3 shows that the algorithm converges quickly to a global minimum indicating the successful parameters estimation due to the good similarity between the reference signal and the modeled signal.

The reconstruction of the fundamental, the harmonic, the interharmonic and the residual components in time and frequency domain for each segment based on the estimated parameters $\omega_0$, $\omega_c$, $c_l$ and $c_k$ according to the expression (6) is shown in the following figures.

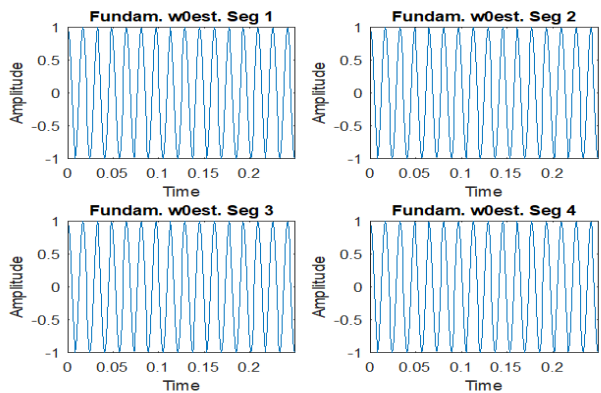

Figure 4: Fundamental component reconstruction in time domain for 4 segments based on the estimated $\omega_0$.

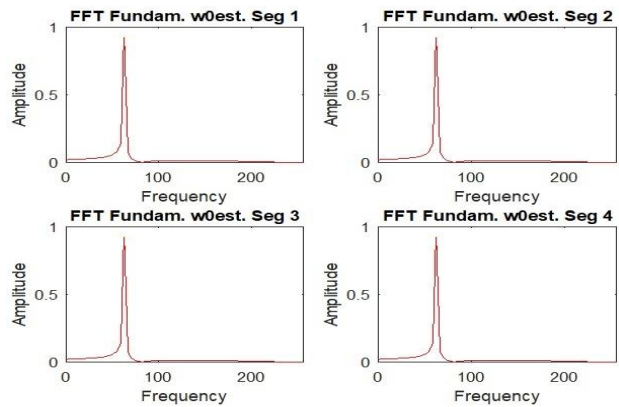

Figure 5: Fundamental component reconstruction in spectral domain for 4 segments based on the estimated $\omega_0$.

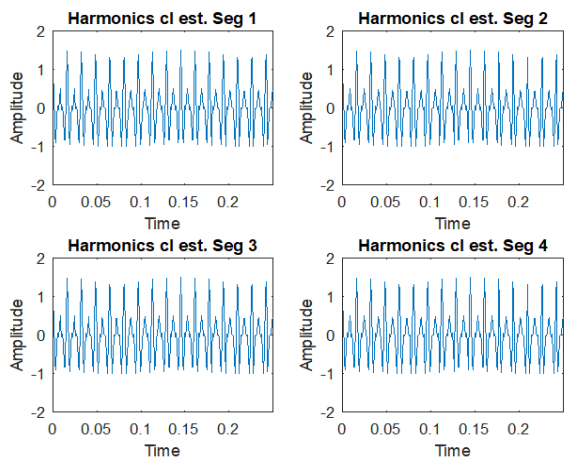

Figure 6: Harmonic component reconstruction in time domain for 4 segments based on the estimated $c_l$.

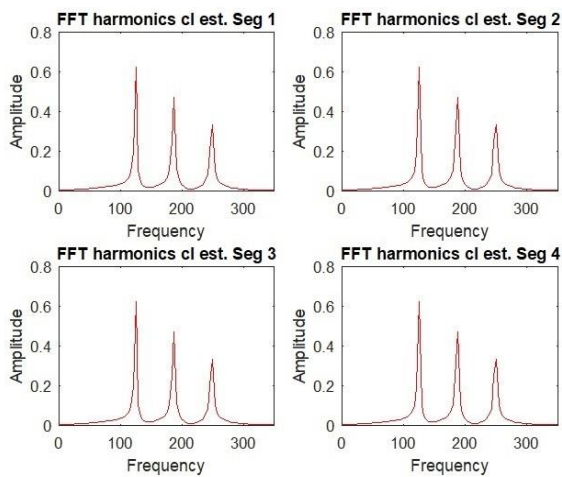

Figure 7: Harmonic component reconstruction in spectral domain for 4 segments based on the estimated $c_l$.

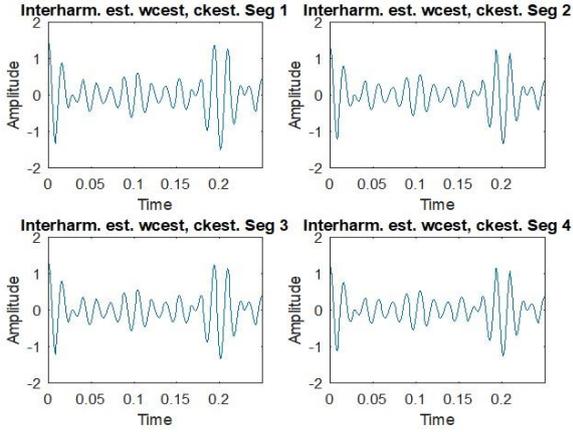

Figure 8: Interharmonic component reconstruction in time domain for 4 segments based on the estimated $c_k$ and $\omega_c$.

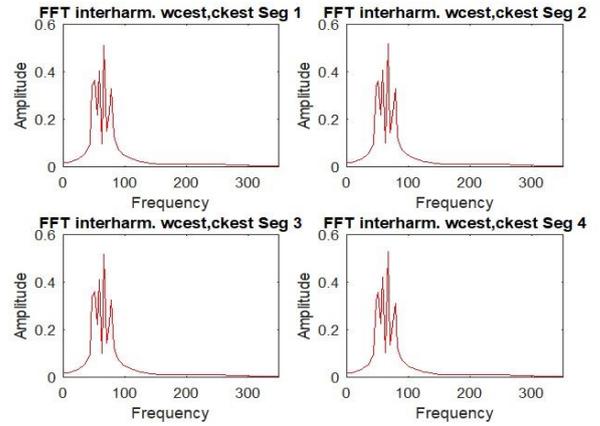

Figure 9: Interharmonic component reconstruction in spectral domain for 4 segments based on the estimated $c_k$ and $\omega_c$.

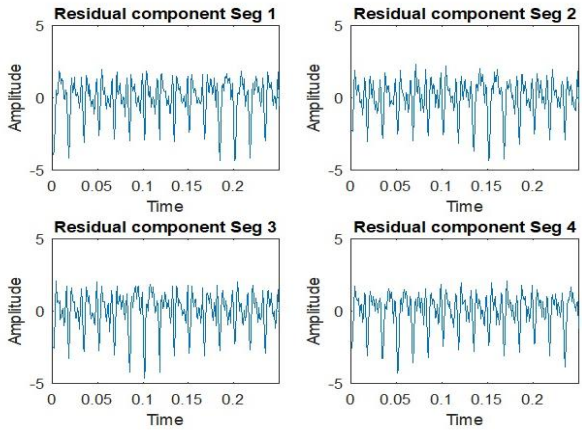

Figure 10: Residual component in time domain for 4 segments based on parameters estimation.

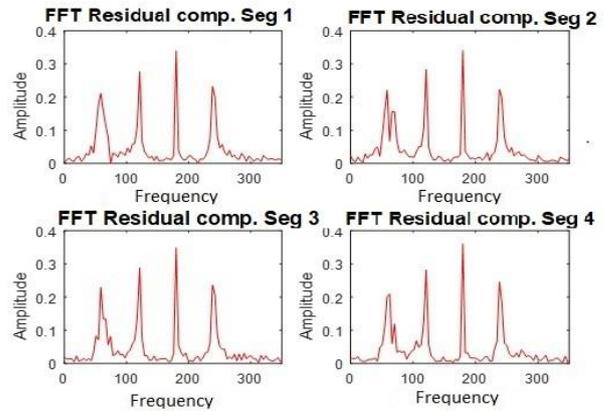

Figure 11: Residual component in time domain for 4 segments based on parameters estimation.

It can be seen that the figures of the reconstructed components are globally similar to the signal theory as detailed in Section 2. Indeed, Fig. 4, Fig. 6 and Fig. 8 show that the signal pattern in time domain is similar to the one found in theory even though we note that there is a very small amplitude difference which varies slightly with the different segments. As for Fig. 5, Fig. 7 and Fig. 9, we can see that most of the frequency spikes of each component are reasonably close to the theorical frequencies.

The difference is the most important for the fundamental component and this can be explained by the fact that it is affected by the predominant fundamental frequency dynamics and by the near presence of the interharmonics. Moreover, the limited number of the signals data points N as well as the associated FFT limitations have an important impact on the component reconstruction accuracy.

Regarding the harmonic component in Fig. 7, we can see that the amplitudes are very close to the theoretical values. As for the interharmonic component in Fig. 9, we observe that the relevant frequency spikes have mostly appropriate values. However, some of the frequency information is hidden. This phenomenon is due to the fact that the estimated frequencies are so close that they overlap when reconstructed since the frequency gap between them is small (5 Hz) especially with the limited number of data points N.

Finally, the residual components in Fig. 10 and Fig. 11 contain the remaining parts of the all the other components since we can see frequency information in the relevant frequencies in addition to noise dynamics. However, the amplitudes of the remaining frequency spikes are distinctly small and they vary very little from a segment to another.

## 5. VALIDATION

In order to evaluate the algorithm's performance and the parameters $\omega_0$, $\omega_c$, $c_l$ and $c_k$ estimation accuracy, Monte Carlo (MC) simulations of the Root Mean Square Error (RMSE) between the estimated signal and the reference signal have been implemented. The RMSE is expressed as:

$$\text{RMSE} = \frac{1}{M}\sqrt{\sum_{j=1}^{M}(y_{est}[j] - y_{ref}[j])^2} \qquad (11)$$

Where M is the observations number, $y_{est}$ represents the estimated signal and $y_{ref}$ represents the reference signal.

The number of carried out MC iterations is $J = 200$ and an iteration represents each set of observations.

In order to validate the proposed method, the MC simulations of this RMSE have been compared to the MC simulations of the RMSE resulting from the Multiple Signal Classification (MUSIC) estimation algorithm. This algorithm estimates the pseudospectrum from the signal using Schmidt's eigenspace analysis method. The algorithm performs eigenspace analysis of the signal's correlation matrix to estimate the signal's frequency content. This algorithm is particularly suitable for signals that are the sum of sinusoids with additive white Gaussian noise [17]. The amplitudes and phases associated to the frequencies estimated by the MUSIC algorithm are estimated using Least Square Method (Matlab's Least Square Fitting Toolbox). The results of this comparison for the parameters $\omega_0$, $\omega_c$, $c_l$ and $c_k$ through each of the four signal segments are shown in the following figures.

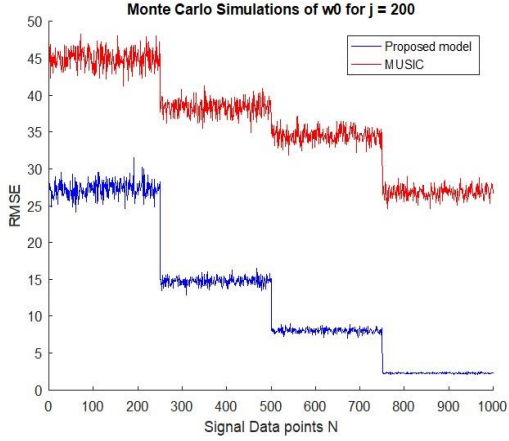

Figure 12: MC simulations of $\omega_0$ by the model and MUSIC

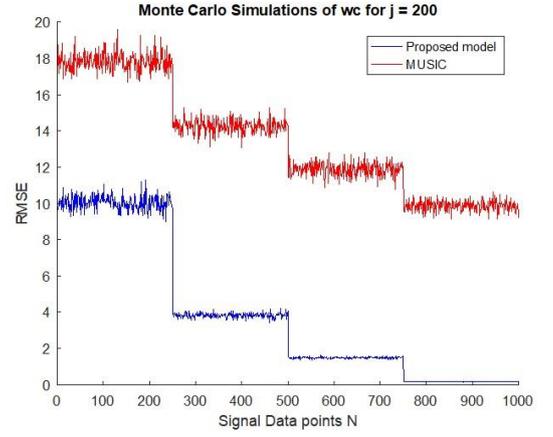

Figure 13: MC simulations of $\omega_c$ by the model and MUSIC

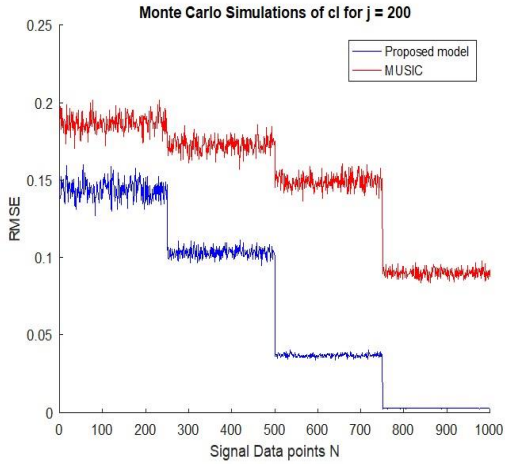

Figure 14: MC simulations of $c_l$ by the model and MUSIC

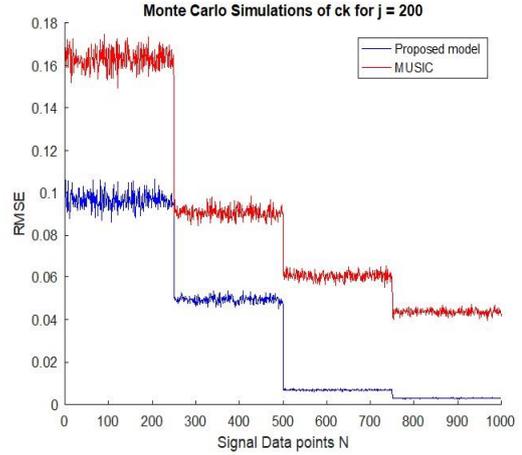

Figure 16: MC simulations of $c_k$ by the model and MUSIC

It can be seen that the MC simulations of the RMSE of the proposed model is globally lower than that of the MUSIC algorithm for all the parameters and it is comparatively more stable especially in the last segments. We can also see in Fig. 12 that the MC simulations of the fundamental angular frequency parameter $\omega_0$ confirm the observations previously detailed in section 5 concerning the notable difference between the estimated value and the theorical value due to increased estimation error. The RMSE is indeed relatively important compared to the RMSE of the other parameters. This can be eventually resolved in the future by specifically adjusting the iterations number I, the learning rate μ of the proposed algorithm as well as the number of data points N. On the other hand, the RMSE values of the rest of the parameters $\omega_c$, $c_l$ and $c_k$ are comparatively low and globally acceptable. Note that compared to the rest of the segments, segment 1 is often the one with the highest RMSE level. This is probably caused by the richer information quantity contained within and this induces a more important generalization error. This explanation is confirmed by the fact that segment 4 is often the one with the lowest error level.

## 6. CONCLUSION

In this paper, the parametric tracking of electrical currents using Gradient Descent algorithm has been implemented. In the context of condition monitoring based on MCSA, we proposed an alternative estimation focused on specific current signal components using a model based on the Vandermonde matrix. This model highlights four major parameters derived from the current components that are relevant for fault monitoring. These parameters have been estimated using the Gradient Descent algorithm and the tracking of the reconstructed current components in the time and the spectral domains based on the estimated parameters has been successfully implemented. The evaluation results show that the estimation of the interharmonic components is accurate. However, the fundamental and harmonic components errors are relatively more important but since these components are less relevant for fault detection, the results are globally promising. Nonetheless, there is still room for future improvements in regards to enhancing the algorithm capabilities with advanced optimization techniques especially in regards to the choice of the learning rate and the iterations numbers, as well as using additional reference signals of different nature and in different configurations in order to assess the algorithm's robustness.

## REFERENCES

[1] S. Choi, E. Pazouki, J. Baek and H. R. Bahrami, "Iterative Condition Monitoring and Fault Diagnosis Scheme of Electric Motor for Harsh Industrial Application," in IEEE Transactions on Industrial Electronics, vol. 62, no. 3, pp. 1760-1769, 2015.

[2] D. Miljkovic, "Brief Review of Motor Current Signature Analysis", CrSNDT Journal, vol. 15, pp.15-26, 2015.

[3] A. S. Fontes, C. A. V. Cardoso and L. P. B. Oliveira, "Comparison of techniques based on current signature analysis to fault detection and diagnosis in induction electrical motors," Electrical Engineering Conference (EECon), pp. 74-79, 2016.

[4] M. S. R. Krishna and K. S. Ravi, "Fault diagnosis of induction motor using Motor Current Signature Analysis", International Conference on Circuits, Power and Computing Technologies (ICCPCT), vol. 3, pp. 180-186, 2013.

[5] M. Benbouzid, "A Review of induction motors signature analysis as a medium for fault detection", IEEE Transactions on Industrial Electronics, vol. 47, no. 5, pp. 984–993, 2000.

[6] G. Bracamonte, J. E. Ramirez-Cortes, J. M. de Jesus Rangel-Magdaleno, J., Gomez-Gil and P. Peregrina-Barreto, "An approach on MCSA based fault detection using independent component analysis and neural networks", IEEE Transactions on Instrumentation and Measurement, vol. 68, pp.1353- 1361, 2019.

[7]   S. Singh, A. Kumar and N. Kumar, "Motor Current signature analysis for bearing fault detection in mechanical systems", Procedia Material Science, vol. 6, pp. 171-177, 2014.

[8]   De Jesus Romero-Troncoso, R., "Multirate signal processing to improve FFT-based analysis for detecting faults in induction motors", IEEE Transactions on industrial informatics, vol. 13, no. 3, pp. 1291 – 1300, 2017.

[9]   E. Elbouchikhi, V. Choqueuse, F. Auger and M. E. H. Benbouzid, "Motor Current Signal Analysis Based on a Matched Subspace Detector," in IEEE Transactions on Instrumentation and Measurement, vol. 66, no. 12, pp. 3260-3270, 2017.

[10]   V. Choqueuse, "Apports des techniques de traitement du signal paramétriques pour l'analyse des signaux électriques et les communications optiques cohérentes", Traitement du signal et de l'image [eess.SP] ⟨tel-02980963⟩, Université de Bretagne Occidentale, 2020.

[11]   P. Stoica, and R. Moses, "Spectral Analysis of Signals", Upper Saddle River, NJ: Pearson Prentice Hall, vol. 447, 2005.

[12]   M. Frini, A. Soualhi, M. El Badaoui, "Gear faults diagnosis based on the geometric indicators of electrical signals in three-phase induction motors", Mechanism and Machine Theory, Vol. 138, pp. 1-15, 2019.

[13]   T., Bafitlhile L. Zhijia and L. Qiaoling, "Comparison of Levenberg Marquardt and Conjugate Gradient Descent optimization methods for simulation of streamflow using Artificial Neural Network", Advances in Ecological and Environmental Research, vol. 3, no.11, 2018.

[14]   E. M. Dogo, O. J. Afolabi, N. I. Nwulu, B. Twala and C. O. Aigbavboa, "A Comparative Analysis of Gradient Descent-Based Optimization Algorithms on Convolutional Neural Networks," International Conference on Computational Techniques, Electronics and Mechanical Systems (CTEMS), pp. 92-99, 2018.

[15]   P. Baldi, "Gradient descent learning algorithm overview: a general dynamical systems perspective", IEEE Transactions on Neural Networks, vol. 6, no. 1, pp. 182-195, 1995.

[16]   M.Garrick, D. Jaeger and F.Harris, "Design and application of a Hilbert Transformer in a Digital Receiver", SDR 11 Technical Conference and Product Exposition, pp. 1-7, 2011.

[17]   O. Das, J.S. Abel, J.O. Smith, "Fast MUSIC - An efficient implementation of the MUSIC algorithm for frequency Estimation of approximately periodic signals", Proceedings of the 21st International conference on Digital Audio Effects, pp. 1-7, 2018.